\documentclass[floatfix,aps,pre,showpacs,showkeys,groupedaddress]{revtex4}
\usepackage{graphicx}
\usepackage{amssymb}
\usepackage{amsmath}

\begin{document}

\title{From one solution of a $3$-satisfiability formula to a solution cluster: Frozen variables and entropy}

\author{Kang Li, Hui Ma, and Haijun Zhou}
\affiliation{Institute of Theoretical Physics, Chinese Academy of Sciences, Beijing 100190, China}

\date{\today}
\begin{abstract}
A solution to a $3$-satisfiability ($3$-SAT)  formula can be expanded into a 
cluster, all other solutions of which  
are reachable from this one through a sequence of single-spin flips.
Some variables in the solution cluster are frozen to the same spin values by one of two
different mechanisms: frozen-core formation and long-range frustrations. While frozen cores are
identified by a local whitening algorithm, long-range frustrations are very difficult to trace, and
they make an entropic belief-propagation (BP) algorithm fail to converge. 
For BP to reach a fixed point the spin values of a tiny
fraction of variables (chosen according to the whitening algorithm) are
externally fixed during the iteration. From the calculated entropy values, we infer that, for
a large random $3$-SAT formula with constraint density close to the satisfiability threshold,
the solutions obtained by the survey-propagation or the walksat algorithm belong neither to 
the most dominating clusters of the formula nor to the most abundant clusters.
This work indicates that a single solution cluster of
a random $3$-SAT formula may have further community structures.
\end{abstract}


\maketitle

\section{Introduction}
\label{sec:introduce}

The $K$-satisfiability ($K$-SAT) problem is a prototypical constraint satisfaction problem in the
non-deterministic polynomial complete (NP-complete) complexity class \cite{Cook-1971}. 
Statistical physicists
became interested in this computer science problem since the discovery 
of phase-transition phenomena in the
ensemble of random $3$-SAT formulas in the early $1990$s.  
Randomly generated $3$-SAT formulas were found to
be either almost always satisfiable or almost always unsatisfiable 
depending on whether or not the density of
constraint $\alpha$ [defined by Eq.~(\ref{eq:alpha}) below] is 
lower than a critical value $\alpha_{s}$ (the
SAT-UNSAT transition point) \cite{Mitchell-etal-1992,Kirkpatrick-Selman-1994}. 
Furthermore, random $3$-SAT formulas
whose satisfiability being most difficult to resolve all have constraint 
densities close to the critical value
$\alpha_{\rm s}$ \cite{Mitchell-etal-1992}. A lot of theoretical work (see, e.g.,
Refs.~\cite{Monasson-Zecchina-1996,Biroli-etal-2000,Franz-etal-2001,Cocco-Monasson-2001,Mezard-etal-2002,Mezard-Zecchina-2002,Mezard-etal-2005-a,Mezard-etal-2005,Krzakala-etal-PNAS-2007,Montanari-etal-2008,Zhou-2008}) 
has been done to understand the satisfiability transition in the random $K$-SAT 
problem and the rapid increase of resolution time as
the constraint density $\alpha$ approaches $\alpha_{s}$.

In the SAT phase with constraint density $\alpha$ close
to $\alpha_{s}$,  the solution space of a typical large random $K$-SAT formula ($K\geq 3$) can be
grouped into many clusters. The solution clusters are not homogeneous in size,
some clusters may contain many more solutions than others. Therefore the solution clusters
are characterized by a (continuous or discontinuous) spectrum of entropy densities
 \cite{Montanari-etal-2008,Zhou-2008,Zdeborova-Krzakala-PRE-2007}.  
On the other hand, it is not clear whether different 
solution clusters are separated by high energy barriers or they can be reached one from the other
through paths of low-energy intermediate partial solutions. This is one of the 
major open questions concerning the organization
of the solution space of a random $K$-SAT formulae. 
In this connection, it was recently realized that clustering of the 
solution space in the random $K$-SAT problem does not
pose real difficulty for heuristic local search algorithms 
\cite{Selman-etal-1992,Selman-Kautz-Cohen-1996,Seitz-etal-2005,Alava-etal-2007,Montanari-etal-ARXIV-2007}.
Algorithms such as GSAT, walksat and ChainSAT \cite{Selman-etal-1992,Seitz-etal-2005,Alava-etal-2007} 
appear to be capable of efficiently escaping from valleys in the energy landscape of a 
random $K$-SAT formula. These experimental
experiences led to the conjecture that what really makes finding a satisfying solution 
hard is the presence of frozen variables (see, e.g., Ref.~\cite{Monasson-etal-1999} and more
recent papers \cite{Seitz-etal-2005,Krzakala-Zdeborova-2007,Ardelius-Zdeborova-2008,Zdeborova-Mezard-2008}).
A frozen variable in a
solution cluster is a variable which is the same literal in all the solutions of the cluster.
If a finite fraction of variables are frozen in a given solution cluster, it was argued 
that it would be difficult for a local algorithm to assign values
to all these variables, and that such solutions would be hard to 
find \cite{Montanari-Semerjian-2006,Semerjian-2008}.
The freezing transition
for the random $K$-SAT problem in principle can be estimated by the entropic cavity method of statistical
mechanics \cite{Mezard-Parisi-2001} but extensive
mean-field population dynamics simulations \cite{Montanari-etal-2008,Zhou-2008,Zdeborova-Krzakala-PRE-2007}
 are needed. For the random $3$-SAT problem the known quantitative estimation of
the freezing transition point comes from a finite-size scaling analysis on exact enumeration results
\cite{Ardelius-Zdeborova-2008}.

Earlier mean-field theoretical studies \cite{Krzakala-etal-PNAS-2007,Montanari-etal-2008,Zhou-2008}
on the $K$-SAT problem have focused on ensemble-averaged properties. In this work, we take a
complementary approach and investigate the properties of solution clusters that are associated with
single reference solutions. This study has been driven by two main motivations:
First we wish to know in more
detail the $`$local structures' of the solution space of the random $3$-SAT problem, 
which might be invisible
in ensemble studies; and second we wish to know whether the solutions obtained by the survey-propagation
and the walksat algorithms for a given random $3$-SAT formula are contained in the dominating
 solution clusters
of this formula. Given an initial satisfying
solution  for a $3$-SAT formula, we use the whitening algorithm of Parisi
\cite{Parisi-2002} (see also Refs.~\cite{Braunstein-Zecchina-2004,Seitz-etal-2005})
to determine which variables are frozen (i.e., taking the same spin value) in the associated solution cluster.
A simple mean-field formula [Eq.~(\ref{eq:rho-f})] is also given, which predicts with high precision
the fraction of frozen variables in planted solutions for the type-B random $3$-SAT formulas studied
in this paper. We point out that, even if a reference solution is completely whitable by the whitening
algorithm, some variables
in the associated solution cluster may still be frozen. This is because variable freezing
can be caused by another independent
mechanism, namely long-range frustration among unfrozen variables as discussed in
Refs.~\cite{Zhou-2005a,Zhou-2005b}. When the neighboring unfrozen variables of a variable $i$
 are long-rangely frustrated, this variable very probably will be frozen.
Two heuristic algorithms are constructed to identify variables
that are frozen due to long-range frustrations.
The entropy of the solution cluster associated with a given reference solution is
calculated by the entropic cavity method, taking the reference solution as initial condition 
for the set of
zero-energy belief-propagation (BP) iterative equations
[Eqs.~(\ref{eq:u-iteration}) and (\ref{eq:eta-iteration})]. The
entropy values reported in this paper are consistent with mean-field results of Ref.~\cite{Zhou-2008}.

For large random $3$-SAT formulas with constraint densities close to $\alpha_s$, we find that
if solutions obtained by the survey-propagation (SP)
\cite{Mezard-etal-2002,Mezard-Zecchina-2002,Braunstein-etal-2005},
the walksat \cite{Selman-Kautz-Cohen-1996}, or the belief-propagation-guided decimation
\cite{Krzakala-etal-PNAS-2007,Montanari-etal-ARXIV-2007} algorithm  are used as
initial conditions, the entropic BP algorithm {\em always} fails to reach a fixed point.
Besides the ensemble of completely random $3$-SAT formulas, a set of large random $3$-SAT formulas
containing a pre-specified satisfying solution are also studied, and for each of them several
additional solutions are obtained by the SP
algorithm and the walksat algorithm. For a $3$-SAT formula in this second ensemble,
if the entropic BP iterative equations are
run with the planted solution as the initial condition, a fixed point is quickly reached, but if
a solution obtained by
the SP or the walksat algorithm is used as the initial condition, the iterative equations again 
fails to converge.
This observation suggests that planted solutions and solutions obtained by the SP algorithm are quite
different. In cases when the BP algorithm fail to reach a fixed point, we fix the spin values of
a tiny fraction of variables and then re-run the BP iteration equations. The modified BP
iteration process will converge if this set of externally fixed variables are chosen
 according to the outcome of the whitening program.

In the remaining part of the paper we work exclusively on the random $3$-SAT problem, but the
illustrated approach should be directly applicable to more general cases.
The following section list the ensembles of random $3$-SAT formulas used in this work. In Sec.~\ref{sec:white}
we investigate the whitening algorithm and present a mean-field formula to describe the freezing transition
in a cluster of solutions. And the SpinFlip algorithm and another heuristic search algorithm are 
introduced to search for frozen variables in
a completely white solution. The entropy of the solution cluster associated with a planted solution
is calculated by the entropic BP algorithm in Sec.~\ref{sec:beliefpropagation}.
For solution clusters associated with single SP or walksat solutions,
their entropy values and fraction of frozen variables are calculated in Sec.~\ref{sec:bp-c} by
combining BP with the whitening program. We conclude this work in
Sec.~\ref{sec:conclusion}.

\section{Generation of satisfiable random $3$-SAT formulas}
\label{sec:graph}

\begin{figure}
 \includegraphics[width=0.4\textwidth]{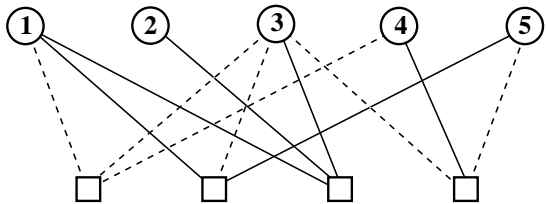}
  \caption{Factor-graph representation for a simple $3$-SAT formula with energy expression $H=(1+\sigma_1)(1+\sigma_3)(1+\sigma_4)/8 +
(1-\sigma_1) (1+\sigma_3) (1-\sigma_5)/8 + (1-\sigma_1) (1-\sigma_2) (1-\sigma_3) / 8 +(1+\sigma_3) (1-\sigma_4) (1+\sigma_5)/8$.}
\label{fig:factor-graph}
\end{figure}

A $K$-SAT formula contains $N$ variables and $M$ constraints (clauses). Each of the $N$ variables ($i, j, k, \ldots$)
has a binary spin state $\sigma_i \in \{ -1, +1 \}$.  Each of the $M$ constraints ($a, b, c, \ldots$)
involves $K$ different variables ($i_a^1, i_a^2, \ldots, i_a^K$) and prohibits these variables from taking a specified 
pattern $(-J_a^1, -J_a^2, \ldots, -J_a^K)$, out of the total number of $2^K$ possible spin patterns of length $K$.
An energy function can be defined for a given $K$-SAT formula as
\begin{equation}
E(\sigma_1, \sigma_2, \ldots, \sigma_N) = \sum\limits_{a=1}^{M} \prod\limits_{i\in \partial a}  \Bigl(\frac{1-J_a^i \sigma_i}{2} \Bigr) \ ,
\label{eq:K-SAT-energy}
\end{equation}
where $\partial a$ means the set of variables involved in constraint $a$.  For a given spin configuration $\vec{\sigma} \equiv
\{ \sigma_1, \sigma_2, \ldots, \sigma_N \}$, the energy $E(\vec{\sigma})$ is equal to the number of unsatisfied clauses.
The zero-energy configurations (if exist) of Eq.~(\ref{eq:K-SAT-energy})
correspond to the solutions of the $K$-SAT formula. A $K$-SAT formula has a convenient factor graph
representation (see the example shown in Fig.~\ref{fig:factor-graph}):
variables are denoted by circular nodes and constraints by square nodes, and there is an edge between
a constraint node $a$ and a variable node $i$ if and only if variable $i$ participates in constraint $a$. 
The edge $(a,i)$ is solid if $J_a^i=1$ and is dashed if $J_a^i=-1$.

In the present paper, we focus on random $K$-SAT formulas with $K=3$.  To generate a random $3$-SAT formula,  $M$ different
triplets $(i, j, k)$ are randomly chosen from the total number of $N (N-1) (N-2)/6$ possible triplets of variable nodes.  A constraint $a$ is applied on each
selected triplet $(i, j, k)$, and it prohibits the simultaneous spin 
assignment $(\sigma_i=-J_a^i ) \wedge (\sigma_j=-J_a^j )\wedge (\sigma_k=-J_a^k )$, where $\wedge$ means  logical
AND. We use two different ways to generate
the prohibited patterns $(-J_a^i, -J_a^j, -J_a^k)$ for the $M$ constraints, which
we call the type-A and type-B formulae (see below). For a given satisfiable $3$-SAT formula
let us denote a particular solution as $\vec{\sigma}^* \equiv \{\sigma_1^*, \sigma_2^*, \ldots, \sigma_N^* \}$.
Since $\vec{\sigma}^*$ is compatible with all the constraints of the formula,
for each constraint $a$ the following associated edge vector
\begin{equation}
\tilde{J}_a \equiv ( J_a^i \sigma_i^*, J_a^j \sigma_j^*, J_a^k \sigma_k^* )\   \hspace*{1.0cm} (i, j, k \in \partial a)
\label{eq:tilde-J-a}
\end{equation}
can have at most two negative elements.  The clauses of the $3$-SAT formula can therefore be grouped into
three types with respective to the reference solution $\vec{\sigma}^*$,
and we denote by $q_0$, $3 q_1$, and $3 q_2$ respectively
the fraction of constraints $a$ whose edge vector $\tilde{J}_a$ have zero, one, and two negative elements. Obviously,
$$
 q_0 + 3 q_1 + 3 q_2 \equiv 1 \ .
$$

For the first ensemble of random formulas used in this paper (type-A formulas), the prohibited spin pattern 
 $(-J_a^i, -J_a^j, -J_a^k)$ of each clause $a$ is independently and uniformly randomly chosen from the total number of eight 
possibilities. Such random formulas are satisfiable with a high probability as long as the
constraint density $\alpha$ defined by
\begin{equation}
 \label{eq:alpha}
 \alpha \equiv M / N
\end{equation}
is less than $4.267$ \cite{Mezard-etal-2002,Mezard-Zecchina-2002}. For each constraint
density $\alpha \in \{ 4.20, 4.21, 4.22, 4.23, 4.24, 4.25\}$, we generate a set of random $3$-SAT formulas of $N=10^6$ variables; for each of
these formulas, we use the survey-propagation algorithm \cite{Mezard-Zecchina-2002,Braunstein-etal-2005} 
(downloaded from Riccardo Zecchina's webpage) to obtain five different
satisfying solutions. 
For $\alpha=4.20$ we also use the walksat algorithm \cite{Selman-Kautz-Cohen-1996} 
(version 45, downloaded from the
walksat homepage) with optimized noise parameter ($p=0.57$ \cite{Seitz-etal-2005}) to obtain another set of solutions.
The solutions serve as initial conditions for the whitening and the belief-propagation 
simulations of the next two sections. For the benefit of later discussions, we refer to 
a solution obtained by the SP algorithm as an SP solution, and a solution obtained by
walksat as a walksat solution.

The second ensemble of satisfiable random formulas (type-B formulas) used in this paper are constructed in
such a way that a pre-given spin configuration $\vec{\sigma}^*$ is a solution. 
Such ensembles with planted solutions have been investigated in the literature
earlier \cite{Barthel-etal-2002}, and are known to have different 
properties from standard random K-satisfiability, see \textit{e.g.}~\cite{Feige-Mossel-Vilenchik-2006,Altarelli-Monasson-Zamponi-2007}.
For each constraint $a$ of the formula, the value of its edge vector $\tilde{J}_a$ as defined by Eq.~(\ref{eq:tilde-J-a}) is
assigned according to the following rule \cite{Barthel-etal-2002}: a uniformly distributed random variable $r \in [0, 1)$ is
first generated; if $r\leq q_0$ then $\tilde{J}_a$ is set to be $(+1, +1, +1)$; if $q_0 < r \leq q_0 + 3 q_1$,
then $\tilde{J}_a$ is chosen uniformly randomly from the set $\{ (+1, +1, -1), (+1, -1, +1), (-1, +1, +1) \}$;
otherwise  $\tilde{J}_a$ is chosen uniformly randomly from the set
$\{ (+1, -1, -1), (-1, +1, -1), (-1, -1, +1) \}$.  For simplicity and without loss of
generality, in this paper we set the
pre-given spin configuration to be $\vec{\sigma}^* = (+1, +1, \ldots, +1)$ when constructing type-B random $3$-SAT formulas.

\section{Freezing of variables in a solution cluster: Two different mechanisms}
\label{sec:white}

Starting from a given solution $\vec{\sigma}^*$ of a satisfiable $3$-SAT formula $F$, 
one can (in principle) build a connected
network of solutions which contains as many solutions of formula $F$ as possible. 
 In this solution cluster, 
two solutions $\vec{\sigma}^1$ and
$\vec{\sigma}^2$ are regarded as being directly connected if and only if 
they differ in the spin value of a single
variable. From one solution of the cluster one can reach any another solution of 
the same cluster by a sequence of
single-spin flips (within the whole solution space of formula $F$). 
We refer to such a connected network of solutions
as a solution cluster (or simply cluster) for formula $F$.
The spin 
states of some variables of the formula may
take the same value in all the solutions of the cluster. Such variables are 
referred to as frozen variables, they 
are strongly constrained in the solution cluster.

\begin{figure}
 \includegraphics[width=0.6\textwidth]{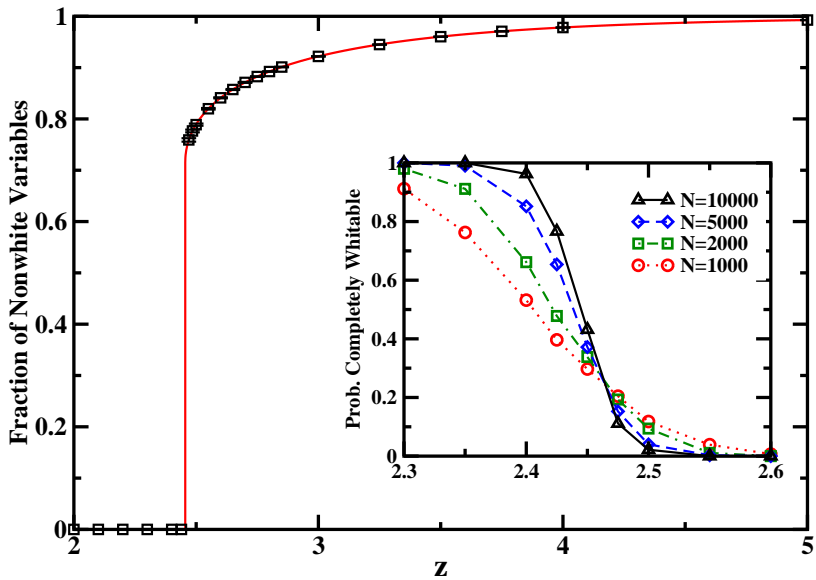}
\caption{\label{fig:scaling}
(Color Online)
The fraction of nonwhite variables in a satisfying solution as a function of the
parameter $z$ as defined in
Eq.~(\ref{eq:z}). The solid line is the mean-field prediction Eq.~(\ref{eq:rho-f}),
 while the square symbols are
the results obtained by averaging over $50$ type-B random $3$-SAT formulas of size $N=10^6$.
  (Inset) The
probability for a satisfiable solution to be
 completely whitable. The data points are obtained by averaging over
more than $1200$ randomly generated type-B $3$-SAT formulas.}
\end{figure}

There are two different mechanisms which cause freezing of variables in a solution cluster.
The first mechanism, which we refer to as 'frozen-core formation', can be described using the
following whitening process \cite{Parisi-2002}. Starting from a given reference solution
$\vec{\sigma}^*$ of a $K$-SAT formula, at step $0$ of the whitening process,
all the clauses $a$ which are simultaneously satisfied by at least two variables of solution
$\vec{\sigma}^*$ are marked as white  and all the variables which do not satisfy any clause or
satisfy only white clauses are marked as white, while the remaining clauses and variables are
all marked as nonwhite. Then at each following step $t$ of the whitening process:
(1) all the nonwhite clauses which are connected to at least one white variables are marked as white; and
then (2) all the nonwhite variables which satisfy only white clauses are marked as white.
The whitening process stops at step $t \geq 1$
if the number of newly whitened clauses (and variables) is
zero.  After this whitening process has finished, if a variable $i$
is left as being nonwhite, one can prove that it is impossible to travel from $\vec{\sigma}^*$ to another
satisfying configuration with $\sigma_i= - \sigma_i^*$ using only satisfying single-spin flips
\cite{frozen-proof}. In other words, the spin of a nonwhite variable node $i$ is frozen to $\sigma_i^*$.
The set of nonwhite variables in the reference solution $\vec{\sigma}^*$ form one or several
frozen-cores. For a variable $i$ in such a frozen-core, there exists at least a clause $a$
of the formula which is satisfied only by variable $i$ in the configuration $\vec{\sigma}^*$
and which is either not connected to other variables or
is connected only to other variables belonging to the same frozen-core of $i$. As the
constraint density $\alpha$ increases, the freezing of variables due to frozen-core formations
 is therefore a phenomenon of bootstrap percolation.

The final result of the whitening process actually is independent of the order in which the variables
are being whitened \cite{Parisi-2002}. It then follows that, in  a solution cluster of a  $K$-SAT problem, 
if one of the solutions can be completely whitened, then all the other solutions are also completely whitable.
To prove this, let us suppose the contrary could be true, i.e., there exist two
solutions of the same cluster, $\vec{\sigma}^1$ and $\vec{\sigma}^2$, with
$\vec{\sigma}^1$ being completely whitable and $\vec{\sigma}^2$ containing a nonempty set $A$
of nonwhite variables.  One can change $\vec{\sigma}^2$ into  $\vec{\sigma}^1$ by following
a path of single-spin flips, and at at each hopping the flipped variable is whitened.
During this transition process no variables of set $A$ is flipped or whitened, as they are frozen
variables.  
After $\vec{\sigma}^1$ is reached, as it is completely whitable,
the remaining  variables (including those in 
set $A$) can all be whitened starting from the partially whitened pattern.
Therefore $\vec{\sigma}^2$ must also be completely whitable and set $A$ should be empty.

We denote the number of nonwhite variables in the solution $\vec{\sigma}^*$ of a random $3$-SAT
formula as $N_{nw}$. If the three types of clauses mentioned in Sec.~\ref{sec:graph}
are randomly distributed in the formula, a very simple equation can be obtained for
the fraction of nonwhite variables $\rho_{nw} \equiv  N_{nw}/ N$.
Consider a randomly chosen variable node $i$.
 This node in its spin state $\sigma_i^*$ is satisfying some clauses, among which $n_i$ clauses are
satisfied only by node $i$. The total number of
clauses in the $3$-SAT formula which are being satisfied by only one variable of the configuration
$\vec{\sigma}^*$ is equal to $ z N$, with $z$ being expressed as
\begin{equation}
 z= 3 q_2 \alpha \ ,
 \label{eq:z}
\end{equation}
where $q_2$ was defined in Sec.~\ref{sec:graph}. Therefore for a large formula with $N \gg 1$, the integer $n_i$ is distributed according to
the Poisson distribution
$\mathbb{P}(n_i) = e^{-z} z^{n_i} /
n_i !$. Variable node $i$ will be nonwhite if, among these
$n_i$ neighboring clauses, there is at least one clause $a$ whose other two connected variable nodes are
both nonwhite. Then the probability of a randomly chosen variable node $i$ being nonwhite is
determined by the following self-consistent equation:
\begin{equation}
\rho_{nw} = 1-\sum\limits_{n_i= 0}^{\infty} \frac{ e^{-z} z^{n_i}}{n_i !} \bigl(1-\rho_{nw}^2\bigr)^{n_i}
 = 1-\exp(-z \rho_{nw}^2 ) \ .
 \label{eq:rho-f}
\end{equation}
For $z$ less than a critical value $z_{nw} = 2.45541$,  Eq.~(\ref{eq:rho-f}) has only the trivial 
solution
$\rho_{nw}= 0$. While for $z > z_{nw}$, another stable positive solution of Eq.~(\ref{eq:rho-f}) appears,
with $\rho_{nw} \geq 0.715332$. The freezing transition at
$z=  2.45541$ is
a first-order bootstrap transition.

Equation (\ref{eq:rho-f}) is confirmed to be valid for planted solutions of type-B random $3$-SAT
formulas (see Fig.~\ref{fig:scaling}). The inset of Fig.~\ref{fig:scaling} shows that,
 when the parameter $z$  defined in Eq.~(\ref{eq:z}) increases slightly
around $2.46$, the probability for the planted solution of a random type-B
formula to be completely whitable drops quickly from $\approx 1$ to $\approx 0$,
 and the slope of this decrease become sharper for larger formulas.
The values of the fraction of nonwhite variables as obtained from these simulations are in 
very good agreement with the mean-field prediction Eq.~(\ref{eq:rho-f}). This good agreement
indicates that the freezing phenomenon in the planted solutions of random type-B formulas can
be completely explained by the formation of frozen cores.

For type-A formulas it is an empirical fact that solutions that can be found efficiently
on large instances using algorithms known today
are always white~\cite{Maneva-etal-2005,Seitz-etal-2005,Achlioptas-RicciTersenghi-2006}
(but see more recent simulation results of Ref.~\cite{Kroc-Sabharwal-Selman-2007}).
We have generated type-A random $3$-SAT formulas with $N=10^6$ and constraint density
$\alpha \in [4.20, 4.25]$, and used the SP algorithm to find solutions to these instances.
For $\alpha=4.2$ we used in addition walksat with noise parameter $0.57$ since this (and other)
stochastic local search heuristics are also known to be effective at these constraint densities
\cite{Seitz-etal-2005,Alava-etal-2007}.
Interestingly, the fractions of constraints satisfied by one, two or three variables appear
to depend only weakly on the constraint density, and is for the solutions found by SP,
$q_0 \approx 0.128$, $q_1 \approx 0.135$, and  $q_2 \approx 0.157$ for all
solutions of instances in this range. The solutions found by walksat at $\alpha=4.2$
display also practically the same values, e.g., $q_2 \approx 0.155$.
These solutions of $N=10^6$ all have a value of $z \approx 2.0$
considerably lower than the critical value $z_{nw}$, and all these solutions are found to be
completely white. 
For a type-A random graph of smaller sizes $N=10^3-10^4$ and constraint density $\alpha=4.2$,
besides finding many completely white solutions,
the walksat algorithm is also able to
reach partially frozen solutions that contain a large fraction of frozen variables, 
if a non-optimal value of
the noise parameter (e.g., $p=0.45$) is used and a long search time is permitted (Lukas Kroc,
private communication). These non-completely white solutions also have a value of $z\approx 2.0$.
The mean-field formula Eq.~(\ref{eq:rho-f}), which does not
consider any correlations in the distributions of the three types of constraints of the
studied $3$-SAT formula, therefore fail to
describe these partially frozen solutions.

\begin{figure}
 \includegraphics[width=0.4\textwidth]{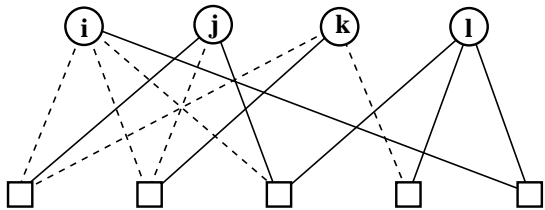}
\caption{\label{fig:frustration}
Frustration effect in a completely white solution $\{ \sigma_i =1,  \sigma_j = 1, \sigma_k = 1, \sigma_l=1 \}$
for a $3$-SAT of $N=4$ variables and $M=5$ constraints. Variable $l$ is frozen to the spin value $\sigma_l = 1$.
}
\end{figure}

Formation of frozen cores is not the only cause of variable
 freezing in the solution clusters of a $K$-SAT formula.
Figure~\ref{fig:frustration} is a very simple example showing that,
even if a solution to a $K$-SAT formula is
completely whitable it can contain frozen variables. In the solution
$\sigma_i= \sigma_j = \sigma_k = \sigma_l = 1$, variable $l$ in satisfying
three white clauses. The two neighbors ($i$ and $k$) of variable $l$ are both
unfrozen variables.  If $l$ is flipped to $\sigma_l = -1$,
variable $i$ should keep $\sigma_i=1$ while
variables $k$ should be flipped to $\sigma_k = -1$. Variable $i$ then
requires variable $j$ to keep the value $\sigma_j=1$, but variable $k$
requires $j$ to be flipped to $\sigma_j=-1$. Such a frustration therefore
prohibited variable $l$ from taking the value $\sigma_l=-1$, although it
is whitable in the whitening process. To speak more generally, with respect
to a solution $\vec{\sigma}^*$ to a $K$-SAT formula, 
let us denote by $\partial_{\sigma_i^*} i$
the set of nearest-neighboring clauses of $i$ which are satisfied by
the spin value $\sigma_i = \sigma_i^*$, and let us assume that each of
these clause $a\in \partial_{\sigma_i^*} i$ can be satisfied in some
solutions of the cluster associated with $\vec{\sigma}^*$ {\em even if
$\sigma_i= - \sigma_i^*$, provided that all  other clauses $b\in \partial_{\sigma_i^*} \backslash a$ 
are removed from the formula}.
Then each clause in the set $\partial_{\sigma_i^*} i$ for sure is whitable in
the solution cluster. But variable $i$ will still be frozen to
the value $\sigma_i^*$ if not all of these clauses can be
{\em simultaneously} satisfied without the need of variable $i$. The 
second mechanism of variable freezing is therefore the closure of
frustration loops.
 In a large random $K$-SAT formula, because the existence of extremely
many loops and because most of  these loops are of length $\log(N)$ or
longer, this freezing mechanism is referred to as
'freezing by long-range frustrations'. 

 Long-range frustrations were analysed
in previous studies \cite{Zhou-2005a,Zhou-2005b} on ensemble of random
networks.  For single solutions of a given random $K$-SAT formula, finding
all the variables which are frozen by long-range frustrations however is
not a trivial task. In contrast to freezing by frozen-core formation,
we are not yet able to construct a polynomial algorithm to identify all the
long-rangely frustrated and frozen variables for a solution of a given
$K$-SAT formula. We leave this challenge to future studies and here
present instead two simple stochastic heuristic algorithms.

The first heuristic search
algorithm (method $1$), SpinFlip, performs a (slightly biased) random walk in the solution
cluster of the formula.
Starting from a solution $\vec{\sigma}^*$ to a $K$-SAT formula $F$, SpinFlip
records and updates the current satisfying configuration of the formula and
three variable sets: 
set $V_1$ contains all the variables that have
already been flipped at least once (unfrozen variables), 
set $V_2$ contains all the variables that have not yet been
flipped and that are currently being flippable, and set $V_3$ contains
all the variables that belong to set $V_1$ and that are currently being flippable. 
In each elementary
trial of the program, if $V_2$ is not empty, a variable
in set $V_2$ is randomly chosen, otherwise a variable in set $V_3$ is
randomly chosen; the spin of this variable is flipped and
the sets $V_1$, $V_2$ and $V_3$ are then updated.
The SpinFlip program is iterated for many steps (each of which consisting
of $N$ consecutive elementary trials) until no new unfrozen variables can
be identified in the last $n$ (say $n=10^6$) consecutive steps.
As SpinFlip is an incomplete algorithm, it may fail to identify some
unfrozen variables of a (completely whitable) solution, 
but we anticipate that most of the unfrozen variables will be discovered
if the program is running for a very long time. 

Besides reporting a set of unfrozen variables, the SpinFlip program can also be used to
explore the fine structure of a solution cluster (paper in preparation). 
The drawback of this random walk algorithm is its slow rate of discovering new unfrozen
variables. The second heuristic search algorithm (method $2$) we used in this work is much faster. This
later algorithm uses information obtained by the whitening program (i.e., $``$to flip variable $i$ you
probably should first flip variables $j$, $k$, $\ldots$''). At each repeat, the algorithm randomly select a
not yet flipped variable and propose a flip. This may cause some of the neighboring clauses (say $a$)
of variable $i$ to be violated. If such a violation happens, then flip a neighboring variable $j$ of
clause $a$, with $j$ being selected according to the causality relationships built by the whitening
program (there are still some freedom in choosing $j$). This flipping process at each branch stops
after a variable  which in its original spin value satisfies all its
neighboring clauses has been flipped. After the whole iteration process stops, 
if no clause is violated then the proposed
spin flip of variable $i$ is accepted and all the flipped variables during this process
 are added to the set of unfrozen  variables.  As a comparison of this program (method $2$) to
the SpinFlip program (method $1$), we notice that, for the example shown in Fig.~\ref{fig:fluctuate},
SpinFlip identifies a total number of $249,923$ unfrozen variables out of $10^6$ variables after
running for $10^6$ steps on a PC (taking about seven weeks), while the whitening-inspired program
is able to identify $265,650$ unfrozen variables in a little bit less than three weeks. A total
number of $228,167$ variables appear in both sets of unfrozen variables. If these two programs are
let to run even longer, more unfrozen variables will be identified, but the rate of finding new
unfrozen variables becomes very low.

\section{The entropic belief-propagation algorithm}
\label{sec:beliefpropagation}

For a given solution $\vec{\sigma}^*$ of a satisfiable $3$-SAT formula, the algorithms mentioned in
the preceding section identify the set of frozen variables in the solution cluster of $\vec{\sigma}^*$.
However, these algorithms do not give information about the spin value
preference of each unfrozen variable node, nor do they estimate the size of the solution cluster.
Now we study in more detail the
statistical property of the solution cluster associated with $\vec{\sigma}^*$
by the cavity method of statistical physics \cite{Mezard-Parisi-2001}.

According to the current statical physics picture, the satisfying solutions of a random $3$-SAT formula
with constraint density $\alpha > 3.86$ are distributed into exponentially many clusters, each of which
in turn contains an exponential number of solutions.
Different solution clusters may have different statistical properties. To characterize such a complex solution
space structure, a  cavity approach which corresponds to 
the mean-field first-step replica-symmetry-breaking spin-glass
theory \cite{Mezard-Parisi-2001} was used in Refs.~\cite{Krzakala-etal-PNAS-2007,Montanari-etal-2008,Zhou-2008}.
In the present paper, as we are interested in single solution clusters of a random $3$-SAT formula, a
replica-symmetric cavity approach is exploited. This cavity approach can be expressed in terms of a set of
belief-propagation (BP) iterative equations (see, e.g., Refs.~\cite{Braunstein-Zecchina-2004,Braunstein-etal-2005,Montanari-etal-2008,Zhou-2008}).

Before we write down the BP equations let us notice however that, the concept of cluster used in this
Section is not strictly equivalent to that defined
in the preceding Section III. In the mean-field spin-glass theory, a cluster (also called macroscopic state)
refers to a sub-space in the system's configuration space which satisfies the so-called clustering property
\cite{Mezard-Parisi-2001}, namely that the spin values of two distantly
separated variable nodes are not correlated. In a macroscopic state, the point-to-set 
correlation between a randomly chosen variable $i$ and variables separated from $i$ by a
shortest-path distance $d$ also decays exponentially with this distance for $d$ large enough
(see Refs.~\cite{Montanari-Semerjian-2006,Krzakala-etal-PNAS-2007} for
more details).
When this clustering property holds, in a given cluster $C$, the joint distribution 
$\mathbb{P}(\sigma_i, \sigma_j, \ldots)$
of spins for a set of distantly separated variables ($i$, $j$, $\ldots$) can
be written in a factorized form:
\begin{equation}
\label{eq:clusteringproperty}
\mathbb{P}(\sigma_i, \sigma_j, \ldots ) = {P}_i(\sigma_i) {P}_j(\sigma_j) \ldots \ , 
\hspace*{1.5cm} (i, j, \ldots {\rm being\; far\; apart})
\end{equation}
where ${P}_i(\sigma_i)$ is the marginal distribution of spin $\sigma_i$ in cluster $C$. Equation (\ref{eq:clusteringproperty})
may not necessarily  be a good approximation for a solution cluster of a satisfiable $3$-SAT formula.
Nevertheless, it turns out that for a large random $3$-SAT formula which has a very sparse factor graph representation,
if the BP iterative algorithm converge to a fixed point,  it always predicts the same set of frozen variables as the whitening
algorithm does. In this case, the BP approach presumably gives an accurate and comprehensive
description of the solution cluster under study.

\subsection{Iterative equations for the entropic belief-propagation algorithm}

In a solution cluster for a $3$-SAT formula $F$, we define
$\eta_i$ as the log-likelihood of variable $i$ to be in the spin-up state, i.e.,
\begin{equation}
 \label{eq:eta}
 \eta_i \equiv \log\Bigl(\frac{P_i (+1) }{P_i(-1)} \Bigr) \ .
\end{equation}
We also define the cavity log-likelihood $\eta_{i\rightarrow a}$ as
\begin{equation}
\label{eq:etaia}
 \eta_{i\rightarrow a} = \log\Bigl( \frac{P_{i\rightarrow a}(+1)}{P_{i\rightarrow a}(-1)} \Bigr) \ ,
\end{equation}
where $P_{i\rightarrow a}(\sigma_i)$ is the probability for variable $i$ to take the spin  $\sigma_i$ if
it is not constrained by clause $a$. We denote by $\exp(u_{a\rightarrow i})$ the fraction of configurations in the
solution cluster in which constraint $a$ is being satisfied by its neighboring variables $j$ other than variable $i$.
Under the assumption that, in the absence of constraint $a$,  the neighboring variable nodes of $a$ are mutually independent of each other,
we can write down the following equation for $u_{a\rightarrow i}$:
\begin{equation}
u_{a\rightarrow i}  = \log\Bigl[1-\prod\limits_{j\in \partial a \backslash i} P_{j\rightarrow a} ( -J_a^j) \Bigr] \ ,
\label{eq:u-iteration}
\end{equation}
where according to Eq.~(\ref{eq:etaia}) $P_{j\rightarrow a}(-J_a^j)$ is related to $\eta_{j\rightarrow a}$ through
\begin{equation}
\label{eq:Piadown}
 P_{j\rightarrow a}(-J_a^j) = \frac{1+J_a^j + (1-J_a^j) e^{\eta_{j\rightarrow a} } }{2 (1+e^{\eta_{j\rightarrow a}})} \ .
\end{equation}
Similarly, if we use again the factorization assumption for the neighboring clauses of a variable node $i$, we get the following equation for $\eta_{i\rightarrow a}$:
\begin{equation}
 \eta_{i \rightarrow a} = \sum\limits_{b\in \partial i \backslash a: J_b^i=-1} u_{b\rightarrow i}
-\sum\limits_{b\in \partial i \backslash a: J_b^i=+1} u_{b\rightarrow i} \ .
\label{eq:eta-iteration}
\end{equation}
In Eqs.~(\ref{eq:u-iteration}) and (\ref{eq:eta-iteration})
$\partial a \backslash i$ means the set of neighboring variables except $i$ for clause $a$, 
and so on for $\partial i \backslash a$.

Equations (\ref{eq:u-iteration}) and (\ref{eq:eta-iteration}) form a set of BP iterative equations, which
were used in various previous studies (see, e.g., Refs.~\cite{Montanari-etal-2008,Zhou-2008}). 
As we are interested in the solution cluster associated with a pre-given solution $\vec{\sigma}^*$, we use the
following initial condition for this set of BP equations.
On each directed edge from a variable node $i$ to a constraint node $a$, at the beginning of the BP  process,
\begin{equation}
\label{eq:eta-initial}
 \eta_{i\rightarrow a} = \left\{ \begin{array}{ll}
                                  + \infty  & \hspace{1.0cm} {\rm if}\;\; \sigma_i^* = +1  \ ,\\
				 - \infty & \hspace{1.0cm} {\rm if} \;\; \sigma_i^* = -1 \ .
                                 \end{array}
\right.
\end{equation}
Starting from this initial condition, the messages $\{ \eta_{i\rightarrow a},
u_{a \rightarrow i} \}$ along all the edges of the factor graph of the $3$-SAT formula are updated according to
Eq.~(\ref{eq:u-iteration}) and Eq.~(\ref{eq:eta-iteration}).
 We have tested a synchronous and a random sequential
BP iteration scheme. In the synchronous updating scheme, in one evolution step, first all the
messages $u_{a \rightarrow i}$ from clauses to variables are updated using Eq.~(\ref{eq:u-iteration}), then
all the messages $\eta_{i\rightarrow a}$ from variables to clauses are updated using Eq.~(\ref{eq:eta-iteration}).
In the random sequential updating scheme, in each evolution step, first a random 
order (say $i_1, i_2, \ldots, i_N$) is made for the $N$ variable
nodes; and for each variable node $i$ in this order, the messages $\eta_{i\rightarrow a}$ (with $a\in \partial i$)
and then $u_{b \rightarrow j}$ (with $b \in \partial i$, $j \in \partial b \backslash i$) are updated.
We have checked that in instances for which the synchronous updating scheme does not drive the messages 
$\{ \eta_{i \rightarrow a}, u_{a\rightarrow i} \}$ to a fixed point, the sequential updating scheme also fails to
do so, and vise versa; while if both the synchronous and the sequential updating schemes lead to convergence of the
iterative equations Eq.~(\ref{eq:u-iteration}) and Eq.~(\ref{eq:eta-iteration}), these two schemes always reach the
same fixed point. This later observation confirms that the BP fixed pints reached by the BP iterative equations are
stable fixed-points. In the numerical simulations, the convergence condition for the BP iteration process is
set to be that the maximal distance $\Delta$ (among all directed edges $a\rightarrow i$ of the
graph) between the values of $e^{u_{a\rightarrow i}}$ obtained in iteration steps $t$ and $t+1$
 is less than
a pre-specified value $5\times 10^{-6}$:
\begin{equation}
	 \Delta = \max\limits_{(a,i)} |e^{u_{a\rightarrow i}(t+1)} - e^{u_{a\rightarrow i}(t)} |
 < 5 \times 10^{-6}  \ .
	\label{eq:fixcriterion}
\end{equation}

After the above-mentioned iteration process has reached a fixed point, the log-likelihood $\eta_i$ for each
variable $i$ as defined by Eq.~(\ref{eq:eta}) can be calculated by
\begin{equation}
 \eta_i = \sum\limits_{a \in \partial i: J_a^i=-1} u_{a\rightarrow i}
-\sum\limits_{a \in \partial i: J_a^i=+1} u_{a\rightarrow i} \ ,
\end{equation}
and the total entropy $S$ of the solution cluster can be estimated by the following equation:
\begin{equation}
\label{eq:entropy}
 S = \sum\limits_{i} \Delta S_i + \sum\limits_{a} \Delta S_a-  \sum\limits_{(i,a)} \Delta S_{ia}  \ .
\end{equation}
In Eq.~(\ref{eq:entropy}) $\Delta S_i$, $\Delta S_a$, and $\Delta S_{ia}$ are,
respectively, the entropy increase due to the addition of variable node $i$, clause node 
$a$, and the edge $(i, a)$ between variable $i$ and clause $a$,  with
\begin{eqnarray}
 \Delta S_i &=&
\log\Bigl[ \exp(\sum\limits_{a\in \partial i: J_a^i=-1} u_{a\rightarrow i} ) + \exp(\sum\limits_{a\in \partial i: J_a^i=+1} u_{a\rightarrow i})  \Bigr]  \ ,  \label{eq:deltaSi}\\
\Delta S_a &=&
\log\Bigl[ 1- \prod\limits_{i\in \partial a} P_{i\rightarrow a}(-J_a^i) \Bigr] \ , \label{eq:deltaSa} \\
\Delta S_{ia} &=&
\log\Bigl[1-(1-e^{u_{a\rightarrow i}}) P_{i\rightarrow a}(-J_a^i) \Bigr] \ . \label{eq:deltaSia}
\end{eqnarray}
Following the work of Chertkov and Chernyak \cite{Chertkov-Chernyak-2006b} it can be shown  that
the entropy expression Eq.~(\ref{eq:entropy}) corresponds to the zeroth-order
term of a loop series for the entropy of the $3$-SAT formula.
For the sparse factor graph of a large random $3$-SAT formula which contains no short loops, higher order terms
in this loop expansion should  not contributed extensively to the total energy of a solution cluster,
and therefore that the entropy density $s \equiv S/N$ as 
obtained by Eq.~(\ref{eq:entropy}) will be exact in the thermodynamic limit of
$N\rightarrow \infty$.

\subsection{Planted solutions as initial conditions for the BP algorithm}

A set of type-B random $3$-SAT formulas of size $N=10^6$ and different constraint densities
$\alpha \geq 4.0$ are constructed, each containing a planted
satisfying solution $\vec{\sigma}^*$ (see Sec.~\ref{sec:graph} for details).
For each of these problem instances, we
run BP as described above and find that the it always reaches
a fixed point starting from the initial condition Eq.~(\ref{eq:eta-initial}). Furthermore, the set of frozen
variables (i.e., variables with $\eta_i = +\infty$ or $\eta_i = - \infty$ at the fixed point) as predicted by the
BP algorithm are always identical to the set of nonwhite variables discovered by
the whitening algorithm of Sec.~\ref{sec:white}. The convergence of the BP algorithm and the
agreement with the whitening algorithm  suggest that the above-mentioned replica-symmetric mean-field cavity theory is
valid and that the planted solution $\vec{\sigma}^*$ can serve as an appropriate initial condition for the
BP algorithm.

\begin{figure}
 \includegraphics[width=0.6\textwidth]{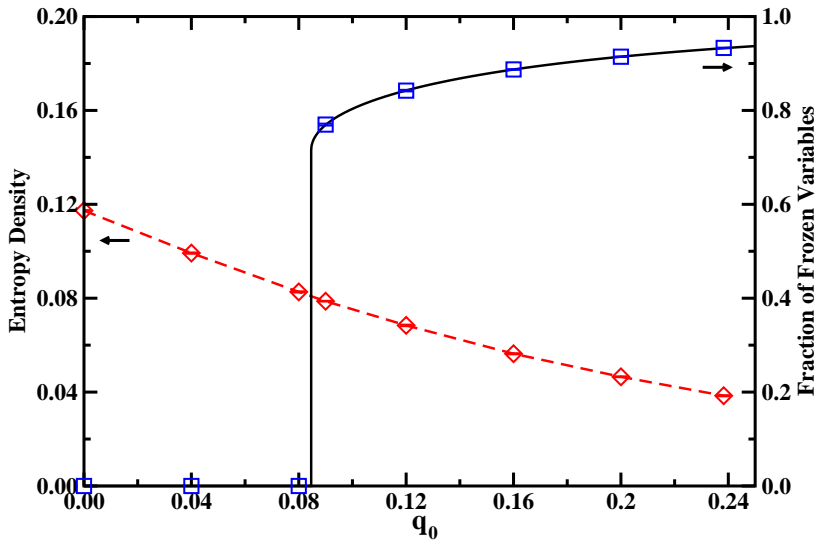}
\caption{ \label{fig:entropy-freezing}
(Color Online) Average entropy density (diamond symbols, dashed 
line being a guide to the eye) and
fraction of frozen variables (square symbols) for $50$ randomly generated
type-B $3$-SAT formulas of $N=10^6$ and $\alpha= 4.2$. The parameters $(q_0, q_1, q_2)$
satisfy Eq.~(\ref{eq:q1q2}).
The solid line is the fraction of nonwhite variables as predicted by Eq.~(\ref{eq:rho-f}).
}
\end{figure}

Figure~\ref{fig:entropy-freezing} shows the BP simulation results for a set of type-B random satisfiable 
$3$-SAT formulas which have $N=10^6$, $\alpha = 4.2$ and
on average equal numbers of initial satisfying and non-satisfying edges. The later restriction is satisfied by
requiring \cite{Barthel-etal-2002}
\begin{equation}
 q_1= (1- 4 q_0)/ 6 \ , \hspace*{1.0cm} q_2 = (1 + 2 q_0)/6 \ ,
\label{eq:q1q2}
\end{equation}
where $q_0$, $q_1$, and $q_2$ are defined in Sec.~\ref{sec:graph}. In this sub-ensemble, the parameter $q_0$ (the fraction
of constraints which are satisfied by three variables in  configuration $\vec{\sigma}^*$) is restricted to
$0 \leq q_0 \leq 0.25$. From Fig.~\ref{fig:entropy-freezing} we know that as $q_0$ increases, the entropy density
$S/ N$ of the solution cluster continuously decreases. For $q_0 \leq 0.08$, there is no frozen variables in
the system (which is consistent with the prediction of Sec.~\ref{sec:white}); while for $q_0 \geq 0.09$, a majority of
the variables are frozen and the fraction of frozen variables is in agreement with the mean-field prediction
Eq.~(\ref{eq:rho-f}). It is interesting to note that at the freezing transition point of $q_0 \approx 0.085$, the
entropy density of the system as a function of $q_0$ does not show any sign of singularity, while the fraction of
frozen variables has a large jump. According to the mean-field cavity theory the entropy densities of solution clusters
in a completely random $3$-SAT formula of $\alpha=4.2$ range from $\approx 0.060$ to $\approx 0.088$ \cite{Zhou-2008},
which are within the range of values shown in Fig.~\ref{fig:entropy-freezing}.

We have applied the whitening-inspired search program (method $2$, see Sec.~\ref{sec:white})
 to the planted solutions of several type-B 
$3$-SAT formulas. In each of these tests, this search program confirms that all the
 white variables of the solution
are nonfrozen variables. In the cluster of a planted solution it appears not to be any long-range
frustrations. This is consistent with the fact that BP always converges with planted solutions.

\section{Entropy of the clusters reached by  survey-propagation and walksat}
\label{sec:bp-c}

We also generate a set of type-A random $3$-SAT formulas of size $N=10^6$ and $\alpha \in [4.20, 4.25]$ and for each of
them, use the survey-propagation algorithm to find a set of satisfying solutions. 
For $\alpha=4.20$ we use in addition walksat as described in Sec.~\ref{sec:graph}.
For $\alpha = 3.90$ and $3.925$ with $N=10^5$, solutions are also obtained by
the belief-propagation-inspired decimation algorithm \cite{Krzakala-etal-PNAS-2007,Montanari-etal-ARXIV-2007}.
The BP algorithms
is then applied on these instances, using these solutions $\vec{\sigma}^*$ as initial conditions. 
We find that for each problem instance, both the synchronous
and the sequential BP schemes predict that there is no frozen variables in the system, consistent with the result of the
whitening algorithm. However, in contrast to the
the preceding subsection, {\em none} of these BP simulations converges to a fixed point
of messages $\{ \eta_{i\rightarrow a}, u_{a\rightarrow i} \}$.
 The messages $\{ \eta_{i \rightarrow a}, u_{a \rightarrow i} \}$
along many edges keeping fluctuating considerably around certain mean values.
Different variables have different amplitude of $\eta$ and $u$ fluctuations. In the random sequential updating
scheme, these fluctuations do not show periodic patterns.

\begin{figure}
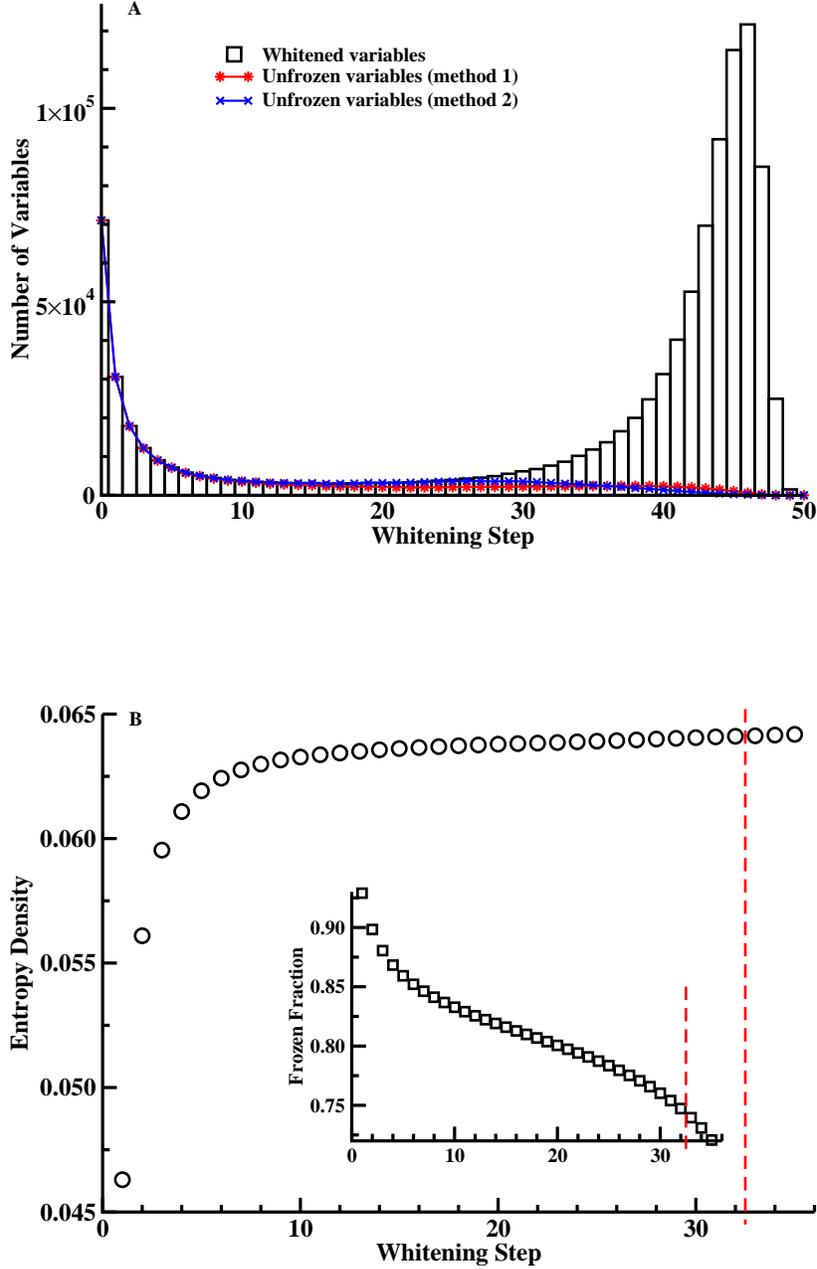

\includegraphics[width=0.6\textwidth]{figure05a.eps}
\vskip 2.0cm
\includegraphics[width=0.6\textwidth]{figure05b.eps}
\caption{\label{fig:fluctuate}
(Color Online)
(a) The $N=10^6$ variables of a random type-A $3$-SAT formula with $M=4.25 \times 10^6$ constraints
is grouped into $51$ sets according to which step $t$ they are whitened by the whitening algorithm.
The SpinFlip algorithm (method $1$) identified $249,923$ variables to be unfrozen after making
$10^6 \times N$ spin flips (in about seven weeks), while the whitening-inspired algorithm (method $2$)
identified $265,650$ variables to be unfrozen after running for about three weeks. The distribution
of these unfrozen variables in each group is shown.
The reference solution is obtained by the SP algorithm.
(b) The entropy density and fraction of frozen variables as predicted by the BP algorithm when the
spin values of all the variables whitened at step $t$ of the whitening process are being externally fixed.
The BP becomes non-convergent at $t \geq 33$ (marked by the dashed lines).
}
\end{figure}

\begin{figure}
\includegraphics[width=0.6\textwidth]{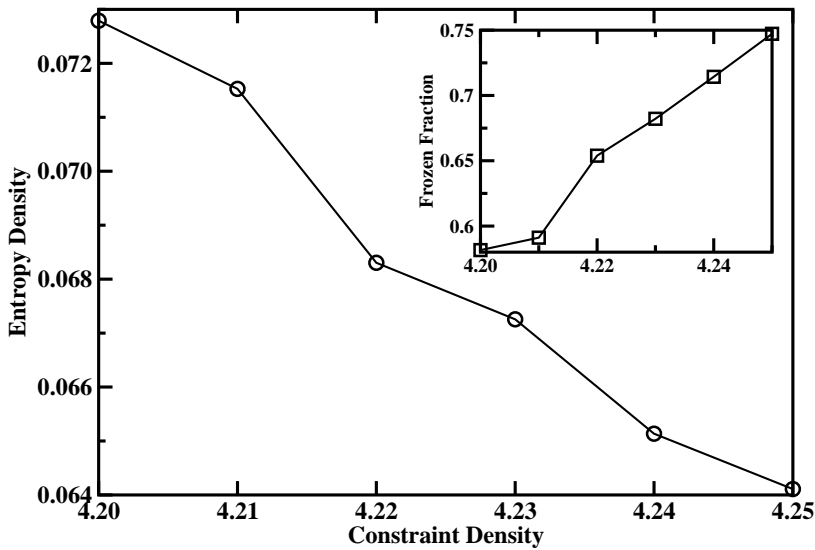}
\caption{\label{fig:entropySP}
The entropy density  of a SP solution-associated solution cluster for a random $3$-SAT formula of size $N=10^6$.
The fraction of frozen variables in the solution cluster as predicted by BP is also shown.
Each data point is obtained by running BP on a single SP-solution by externally fixing a tiny fraction
of variables.
}
\end{figure}

If the non-fixed-point messages  $\{ \eta_{i\rightarrow a}, u_{a\rightarrow i} \}$ are
used to calculate the entropy, Eq.~(\ref{eq:entropy}) reports an entropy density value of
$s \approx 0.090$ at $\alpha= 4.2$, which is equal to the
replica-symmetric entropy density as calculated in earlier studies (see Fig.~2(b) of
Ref.~\cite{Zhou-2008} or Ref.~\cite{Montanari-etal-2008}).
We have further checked that, if the BP iteration starts from the RS initial
condition $ \{ \eta_{i\rightarrow a} =0 \}$, the evolution trajectory
of the messages $(\eta_{j\rightarrow b}, u_{b\rightarrow j})$ on any given edge $(j,b)$ can not be
distinguished from the evolution trajectories starting from a SP- or walksat-solution.
The BP algorithm therefore in some sense forgets its starting point and
does not bring us any cluster-specific information.

The non-convergence of BP is not
due to the different $3$-SAT ensemble used in this section but 
is related with the initial conditions used in the BP. 
To support this claim, we notice that for the type-B random
formulas with $q_0 > 0.01$ studied in Fig.~\ref{fig:entropy-freezing} of the preceding section,
when SP- or walksat-solutions
instead of planted solutions are
used as initial conditions for the BP algorithm, the BP algorithm fails to
converge. At $\alpha=4.2$ and $q_0 > 0.01$, solutions found by SP or walksat are not in the
same cluster as the planted solution. The planted solution correspond to a $`$crystal' phase, while
the SP- and walksat-solutions are all belong to the $`$glassy' phase. 
The set of type-B random formulas with $q_0< 0.01$ used in Fig.~\ref{fig:entropy-freezing} are
in the replica-symmetric phase (only one solution cluster), and then
BP converges to the same fixed-point independent of initial conditions.

As we mentioned in Sec.~\ref{sec:white}, variable freezing can be caused by long-range frustrations
\cite{Zhou-2005a,Zhou-2005b}.  As an example, the SpinFlip algorithm was only able to
confirm that $25$ percent of the variables are unfrozen in a completely white SP-solution for a
type-A $3$-SAT formula with $N=10^6$ variables and constraint density $\alpha=4.25$, even
after running for seven weeks. (The whitening-inspired search algorithm performs a little bit
better, it reported that $26.5$ percent of the variables in
the same solution are unfrozen, after running for about three weeks.) 
As BP uses only local structural information of a graph, it is unable
to detect the globally constrained variables. Some of the variables in the solution can be
externally fixed to make BP converge.
For this exemplar solution,
in Fig.~\ref{fig:fluctuate}A we plot the number of newly whitened variables at
each step of the whitening process, and show how many of them are also
identified as unfrozen variables by the two heuristic programs.
At earlier steps ($t \leq  25$) of the whitening process, most of
the whitened variables are confirmed to be unfrozen variables; but at later whitening steps
most of the newly whitened variables are very difficult to be flipped (if not impossible).
This is expected: as long-range frustration effects are strongly related to the closure
of loops in the graph, only variables which are whitened at steps of order $\log(N)$
(the typical length of a loop) or later can have a large probability of being
frozen or extremely constrained. Since variables that are whitened in the later steps of the
whitening process are extremely difficult to be flipped, they must be extremely constrained in
the reference solution $\vec{\sigma}^*$. Figure~\ref{fig:fluctuate}A then suggests that
a way to make BP converge is to externally
fix the spin values of variables which are newly whitened in certain whitening step (say $t$). If
these variables are fixed, the whitening program can not proceed to the next step $t+1$, and BP, which
is capably of detecting frozen-cores, will then predicts all the variables correspond 
to steps $\geq t+1$
in Fig.~\ref{fig:fluctuate}A to be frozen.

Figure~\ref{fig:fluctuate}B shows the results of the BP after externally fixing a group of
variables. When the spin values of the variables corresponding to whitening step $t$ are all
externally fixed, the BP iteration converges to a fixed point as long as $t \leq 32$. The
calculated entropy density value increases with $t$ only very slowly and the predicted
number of frozen variables decreases with $t$.  BP fails to converge when the
externally fixed variables are from whitening step $t\geq 33$ (the parameter
$\Delta$ as defined in Eq.~(\ref{eq:fixcriterion}) is about $0.5$ even after $2\times 10^4$
iteration steps). Therefore  we take the value of $s = 0.0641$ obtained at $t=32$ as
the entropy density of the solution cluster. The fraction of frozen variables is predicted to be
$0.7472$ at $t=32$, while from the result of the two heuristic search programs (see Fig.~\ref{fig:fluctuate}A)
we know that the real fraction of frozen variables
should be at most $0.713$.
It appears that more variables should be in the frozen state for BP to converge: if we run the BP
iteration by externally fixing all the frozen variables reported by
either of the two search programs, BP again fails to converge. 
It is tempting for us to interpret this observation as follows: The solution cluster can be divided into
many  sub-clusters; in each sub-cluster, besides those variables that are frozen in all the sub-clusters,
some additional variables are frozen.  For a very large random $K$-SAT formula, as there are exponentially
many different combinatorial ways to choose these additional sub-cluster specific frozen variables,
probably the number of sub-clusters is also exponential. These sub-clusters are not separated with each other
by any energy barriers, they are formed by correlation properties.
This conjectured further structural organization
in a single solution cluster will be checked by a future investigation.

We have applied this combined BP and whitening approach to a set of other type-A random $3$-SAT formulas and
SP-solutions and walksat-solutions. 
Some of the results are reported in Fig.~\ref{fig:entropySP}. At $\alpha=4.20$, the entropy densities of
the solution clusters reached by SP and walksat are comparable, being
$s\approx 0.0725$ and $s\approx 0.0715$, respectively. In comparison, the dominant clusters of the
random $3$-SAT problem has entropy density $s\approx 0.088$ and the most abundant clusters have
entropy density $s \approx 0.060$ \cite{Zhou-2008}. At $\alpha = 4.25$, the clusters reached by
SP have entropy density $s\approx 0.064$, while the dominating and most abundant clusters have entropy
density values $s\approx 0.068$ and $s \approx 0.060$, respectively \cite{Zhou-2008}.
It appears that solutions reached by SP and walksat belong neither to one of the dominating clusters or to 
one of the most abundant clusters (the same observation was obtained in Ref.~\cite{DallAsta-etal-2008} for
the random bi-coloring problem). 

We close this Section by mentioning that another way of making BP converge to a fixed point is
by damping \cite{Pretti-2005}. This strategy was not used in the present work because we were not
yet very clear of the physical meaning of the converged fixed-point of the damped BP algorithm
\cite{Pretti-2005}. As explained in this Section,  the action of externally fixing some
variables to make BP converge is physically reasonable, as most of the predicted frozen variables by
the resulting BP fixed-point are highly constrained
variables and very difficult to be flipped (see Fig.~\ref{fig:fluctuate}A). 
Explicitly fixing these variables helps to
remove possible long-range correlations within the solution cluster of the $3$-SAT formula.
Figure~\ref{fig:fluctuate}B demonstrates that the calculated entropy value is not sensitive
to the set of variables that are being externally fixed.
Another different strategy was used in Ref.~\cite{DallAsta-etal-2008} to calculate
the entropy of a single solution cluster.
 
\section{Conclusion}
\label{sec:conclusion}

This paper have studied the statistical property of solution clusters that are associated with
single solutions of a random $3$-SAT formula. It was pointed out that there are two different
mechanisms for the freezing of variables in a solution cluster. Variables which are
frozen due to frozen-core formation can easily be identified
by both the whitening  and the belief-propagation
algorithm. But variables which are frozen due to long-range frustrations can be very difficult to
be identified, as long-range frustrations  involve global and
topological property of the $3$-SAT formula. 
A heuristic search algorithm, SpinFlip, was constructed to search for such variables.

When long-range frustrations exist in a solution $\vec{\sigma}^*$ and the associated cluster of 
random $3$-SAT formula, the BP iteration process starting from the initial condition $\vec{\sigma}^*$
is unable to reach a fixed-point. To overcome this difficult, a tiny set of variables (chosen with the
help of the whitening algorithm) was externally fixed to their initial spin values during the BP iteration. 
When this modified BP process reaches a fixed point, the entropy densities of the solution cluster was
evaluated. It was found that at $4.2 \leq \alpha \leq 4.25$, the solutions obtained by SP or walksat for
a given random $3$-SAT problem are in medium-sized clusters, their entropy densities are
higher than the entropy density of the most abundant clusters in the
formula but lower than the entropy density of the most dominating clusters.

The present work  indicates that at constraint density $\alpha$ close to the satisfiability
threshold, a single solution cluster of a random $3$-SAT formula can be further divided into
sub-clusters. Such further structural organizations, if exist, may be described using the
first-step RSB spin-glass cavity theory \cite{Mezard-Parisi-2001,Mezard-etal-2002}. Further work
along this line will be reported in another paper.

\section*{Acknowledgment}

We thank  Erik Aurell, Lukas Kroc, Jie Zhou for helpful discussions, and
thank Lukas Kroc for sharing his walksat simulation data. We are
grateful to Erik Aurell and Lenka Zdeborov{\'{a}} for their 
critical comments and to Erik Aurell for revising earlier versions of the manuscript.
HZ benefited from the KITPC program ``Collective Dynamics in Information Systems'' (Beijing, Kavli Institute for Theoretical
Physics China, March 01--April 15, 2008) and the NORDITA program `Physics of Information Systems'' (Stockholm, Nordic Institute for
 Theoretical Physics, May 05--May 31, 2008). The kind hospitality of KITPC and NORDITA was gratefully acknowledged by HZ. 
This work was also partially supported by NSFC (Grant No. 10774150).


\begin{thebibliography}{42}
\bibitem[{fro()}]{frozen-proof}
\bibinfo{note}{Let us assume the contrary could be true, i.e., for a non-white
  variable $i$ there exists a path of single-spin flipps from the initial
  configuration $\vec{\sigma}^*$ to another configuration $\vec{\sigma}^\prime$
  which as $\sigma_i^\prime = - \sigma_i^*$. Along this path a total number of
  $m+1$ distinctive variables are flipped at least once. Denote this set of
  variables as $\{j_1, j_2, \cdots, j_m, i\}$ in the temporal order of being
  flipped for the first time. It is easy to show these $m+1$ variables can be
  whitened consecutively. Then we arrive at the conclusion that variable $i$ is
  whitable, which is in conflict with our assumption that $i$ is non-white.
  Therefore, a non-white variable must be a frozen variable.}
\end{thebibliography}
\end{document}